# The standard coil or globule phases cannot describe the denatured state of structured proteins and intrinsically disordered proteins

Francesco Righini[1], Gregory Potel[2], Riccardo Capelli[3] and Guido Tiana[1,4*]

[1]Department of Physics, Università degli Studi di Milano, via Celoria 16, 20133 Milano, Italy
[2]Lawrence Livermore National Laboratory, USA
[3]Department of Biosciences, Università degli Studi di Milano, via Celoria 26, 20133 Milano, Italy
[4]INFN, via Celoria 16, 20133 Milano, Italy

**Abstract**

The concepts of globule and random coil were developed to describe the phases of homopolymers and then used to characterize the denatured state of structured cytosolic proteins and intrinsically disordered proteins. Using multi-scale molecular dynamics simulations, we were able to explore the conformational space of the disordered conformations of both types of protein under biological conditions in an affordable amount of computational time. By studying the size of the protein and the density correlations in space, we conclude that the standard phases of homopolymers and the tools to detect them cannot be applied straightforwardly to proteins.

**Introduction**

It is generally accepted that the denatured state of structured proteins under biological conditions is globular[1], whereas intrinsically disordered proteins (IDPs) are extended coils[2]. According to the hydrophobic-collapse model, the denatured state of structured proteins becomes compact due to the effective attraction between hydrophobic residues[3]. Under biological or mildly denaturing conditions, its volume is only ~30% larger of that of the native state[1] and can expand considerably only when chemical denaturants are added[4,5]. On the other hand, IDPs remain in the coil phase due to the low degree of hydrophobicity and the strong repulsion between charges of the same sign[6].

These conclusions are drawn from the study of the size of proteins in solution. Unlike IDPs, the experimental analysis of the denatured state of structured proteins under biological conditions is not an easy task because of its metastability. When it is stabilized by chemical denaturants it is always reported as an extended coil, with a large radius of gyration $R_g$ and the scaling behaviour $R_g \sim N^{3/5}$ predicted by polymer theory for coils, measured either by intrinsic viscosity[7] or by small angle X-ray scattering[8]. Unfortunately, the same techniques cannot be usually applied to the metastable denatured state under biological conditions, which is the state most relevant to folding (and misfolding). Only for particularly unstable proteins, such as the Δ135-149 mutant of staphylococcal nuclease[9] or the N-terminal SH3 domain of drk[10], it is possible to characterize the denatured state in absence of chemical denaturant. In both cases, the mean size of the denatured state is smaller than that of an ideal chain. Anyway, several studies put the changes of size within the disordered state of proteins under the framework of a coil-globule transition[5,11,12].

Analysis of the radius of gyration alone can often give misleading results as to the equilibrium phase of a polymer. The need for an accurate Kuhn length and the effect of pretranslational swelling[13] make it difficult to use the numerical value of $R_g$ under given conditions to identify the polymeric phase of a protein. Even the scaling behaviour of the radius of gyration with chain length can lead to incorrect conclusions[14].

---

* Correspondence to guido.tiana@unimi.it

The real difference between globular and extended coil phases lies in the density correlations and their fluctuations[15]. Although more difficult to study than the radius of gyration, these properties better describe the physical phase of a polymer at equilibrium.

Anyway, the transition from random globule to random coil is a phenomenon that has been observed and modelled for long homopolymers, whose local properties are invariant to translation along the chain, and cannot be acritically applied to proteins that are stabilised by heterogeneous interactions. In the case of random heteropolymers, mean-field models have shown that the phase diagram is richer than that of homopolymers[16]. However, such analytical models cannot describe phenomena that depend on the specific sequence of amino acids.

Conversely, molecular dynamics simulations can be useful to study the conformational properties of the denatured state of structured proteins and to compare them with those of IDPs. The availability of force fields such as Amber99SB-disp[17], which can recapitulate the conformational properties reported in experiments for both structured and disordered proteins has paved the way for the sampling of non-equilibrium states that escape experimental characterization. The main problem that remains is the computational cost of simulating a polymer that can fluctuate to quite large radii of gyration, thus requiring a large box filled with a large number of water molecules. To mitigate this problem we have designed and applied a multiscale approach that allows us to simulate more compact conformations at the atomic scale with the Amber force field in a smaller box, while more extended conformations are modelled in a coarse-grained manner with Calvados, a $C_\alpha$-model in implicit solvent trained to reproduce small-angle X-ray scattering and paramagnetic relaxation enhancement data[18]. The absence of explicit solvent molecules makes the choice of the size of the simulation box irrelevant and the coarse-grained description of the protein makes the calculations particularly fast. Of course, this model is not able to reproduce the fine-grained structure present in native-like states, such as secondary structures. However, we expect extended conformations to be less dependent on the detailed chemistry of the atomic interactions, and therefore their description can rely on a less fine-grained model. Calvados has been shown to be effective in describing the size of the ensemble of conformations of disordered proteins at different values of ionic strength; however, its coarse-grained nature prevents it from describing the secondary structures that populate transiently the disordered states of proteins in biological conditions[10].

We compared the simulations of the denatured state of a structured protein with the conformational ensemble of an IDP. They are a monomeric variant of the HIV-1 protease (HIV-1-PRΔ96-99) and an unstructured fragment of the Breast Cancer type 1 protein (Brca1-304-394), respectively. HIV-1-PRΔ96-99 folds in tens of seconds in buffer[19], therefore its denatured state is a long-lived metastable state in biological conditions. In both systems, we have approximately the same number of residues and the typical composition of structured proteins and IDP, respectively; the former is rich of hydrophobic residues, the latter is depleted of them and rich in charged amino acids (Fig. S1 in the Supp. Mat.). Since it is known that the ionic strength of the solution has an important effect on the size of IDPs[20], we simulated the dynamics of the two proteins both in a buffer with realistic ionic strength and compared the results with simulations in pure water.

We investigated the geometric properties of the disordered state of these two proteins in with the properties associated with the density correlations. In the case of HIV-1-PRΔ96-99, we focus only on its denatured state, ignoring the fact that eventually it will fold to a globular native state.

In the "Methodology" section we present the multiscale approach that allowed us to sample the conformational space of the two proteins, including the more extended conformations, in a reasonable amount of computer time. Then we show in the "Results" section that both proteins show features of a globular phase, such as their gyration radius and the solvent accessibility of the hydrophobic residues, in water and move towards an ideal-chain behavior in buffer. However, we show that the density correlations and fluctuations are similar in the two solvents and are inconsistent with a globular phase. The reason for that is the presence of transient structure, specific for each protein, which we can detect and which affects its size and its correlation properties.

**Methodology**

*The protein model*
We performed extensive replica exchange simulations of HIV-1-PRΔ96-99 and Brca1-304-394 (Table S1 in the Supp. Mat.) in a realistic buffer using a multiscale approach, where the more compact conformations are simulated in explicit solvent, while extended conformations are simulated with a coarse-grained model. We also compared these simulations with similar simulations performed in water to investigate the effect of the buffer. The simulation of HIV-1-PRΔ96-99 in water is taken from ref. [21], while the others were performed *de novo*. The force field is Amber99SB-disp[17] and the water is described by a tip4p model[22].

*The multiscale approach*
In contrast to other multiscale approaches, here we performed two independent sets of simulations with the atomistic and coarse-grained models, respectively, and only merged the sampled conformations at the end. Since we did not change the parameters of either model, we can rely on their validations as reported in the literature by years of use. Atomistic simulations are performed with a replica-exchange protocol to prevent kinetic trapping that could slow down the equilibration of the system, while simulations with the coarse-grained model are less sensitive to this problem.
We performed the high-detailed atomistic simulations with Gromacs-2022.5[23] patched with Plumed 2.10[24]. We add to the molecular force field a confining harmonic potential to constrain the size of the protein

$$U_{conf} = \theta(d_m - d_0)\, k_{conf}\, (d_m - d_0)^2, \tag{1}$$

where θ is a step function that makes $U_{conf}$ different from zero only when $d_m > d_0$, $d_m$ is the maximum distance between atoms of the protein

$$d_m = b \log \sum_{i,j} \exp\left(\frac{|r_i - r_j|}{b}\right), \tag{2}$$

Where b is a parameter set to 0.05 to make $d_m$ a smooth approximation of the maximum interatomic distance $\max_{i,j}(|r_i - r_j|)$. This potential constrains the polymer to keep the simulation box small and the statistical weights of the sampled conformations are rescaled *a posteriori*, as described below.
Atomistic simulations are performed using 84 replicas with temperatures ranging from 300 K to 420 K. We used a dodecahedral box of diameter 8.5 nm, containing approximately 14,000 water molecules. We chose $k_{conf} = 83.76$ kJ/mol/nm² and $d_0 = 7.3$ nm, which is the maximum distance that prevents the atoms of the protein to stay at a distance of 1.2 nm from each other across the periodic boundary conditions (avoiding self-interaction with short-range non-bonded interactions (using the Gromacs parameters *rcoulomb* = *rdw* = *rlist*=1.0 nm, using PME). HIV-1-PRΔ96-99 is first unfolded at 700K. A preliminary equilibration of 30 ns was carried out for both proteins. To simulate the proteins in buffer, we added to the solution Na$^+$ and Cl$^-$ atoms as listed in Table 1, corresponding to a concentration of 0.150 M.

We recorded conformations every 200 ps, corresponding to the decorrelation time between the conformations of the proteins, estimated from their mutual RMSD (Fig. S2 in the Supp. Mat.). In a total simulation time of the order of 10 ms, the replicas efficiently enhanced the phase space exploration, and the main quantities of interest seem to have reached convergence (Fig. S3-S4 in the Supp. Mat.).

|  | Duration | Cl⁻ ions | Na⁺ ions |
| --- | --- | --- | --- |
| HIV-1-PRΔ96-99 buffer | 20 μs | 43 | 39 |
| Brca1-304-394 buffer | 14 μs | 41 | 39 |
| Brca1-304-394 water | 12 μs | 2 | 0 |
| HIV-1-PRΔ96-99 water | 68 μs | 4 | 0 |

Table 1: Details of the replica-exchange simulations. The simulation of HIV-1-PRΔ96-99 in water is reported in ref. [21].

To describe the conformations with $d_m \gg d_0$, which cannot be sampled by the atomistic simulation due to the effect of $U_{conf}$, we made use of the coarse-grained description provided by Calvados [18]. For each temperature, we performed a 100 ns simulation for each of the two proteins, setting the ionic strength either to 0 or to 0.15M, and recording 2000 conformations. The Amber and Calvados simulations are carried out independently on each other.

The distributions of the visited conformations with the two methods $p_{amb}(r)$ and $p_{cal}(r)$ are the number of occurrences of conformation r encountered in the Amber and Calvados simulations, respectively:

$$p_{amb}(r) = \frac{1}{N_{amb}} \sum_{\bar{r} \in amb} \delta(r - \bar{r}), \qquad p_{cal}(r) = \frac{1}{N_{cal}} \sum_{\bar{r} \in cal} \delta(r - \bar{r}). \tag{3}$$

We assigned a statistical weight to each visited conformation,

$$p(r) = \mathcal{N} \left[ p_{amb}(r)\, w(r)\, \alpha(d_m(r)) + p_{cal}(r)\, \kappa\, [1 - \alpha(d_m(r))] \right] \tag{4}$$

where $\mathcal{N}$ is a normalization constant,

$$w(r) = \frac{e^{\beta U_{conf}(d_m(r))}}{\left\langle e^{\beta U_{conf}(d_m(r))} \right\rangle_{amb}} \tag{5}$$

is the scaling factor that reweights the conformations visited by the biased Amber simulation,

$$\kappa = \frac{\int_{bin} dr\, \frac{e^{\beta U_{conf}(d_m(r))}}{\left\langle e^{\beta U_{conf}(d_m(r))} \right\rangle_{amb}} p_{amb}(r)}{\int_{bin} dr\, p_{cal}(r)} \tag{6}$$

is the scaling factor that matches the distribution of $d_m$ for Amber and Calvados in a particular bin of values of $d_m$ and

$$\alpha(d_m) = \begin{cases} 1 & , d_m < d_0 \\ \frac{d_{max} - d_m}{d_{max} - d_0} & , d_0 \leq d_m \leq d_{max} \\ 0 & , d_m > d_{max} \end{cases} \tag{7}$$

where $d_{max}$ is the largest $d_m$ encountered in the Amber simulation, expresses the relative weight of Amber conformations respect to Calvados conformations depending on $d_m$. In this way, if $d_m$ is smaller than $d_0$ the probability of a conformation is the frequency of being observed in the Amber simulation, according to the ergodic theorem; if $d_m$ is greater than $d_{max}$ only Calvados conformations are taken in account; in the intermediate region, we consider both types of conformations, properly weighted to account for the fact that extended conformations cannot be reached in the Amber simulation because of the constraint potential. The distribution of $d_m$ generated by Calvados is compatible with that of Amber in the range where both can be calculated (Fig. S5 in the Supp. Mat.).

All quantities of interest can be obtained from the p(r) estimated from Eq. (4). For example, the distribution of radius of gyration $R_g$ is

$$p(R_g) = \int dr\, p(r)\delta[R_g(r) - R_g], \tag{8}$$

that can be estimated operatively binning the range of values of $R_g$ and associating to each bin the sum of the statistical weights of all visited conformations, evaluated from Eq. (4).

*Calculation of correlation functions*

An important quantity to evaluate in what phase is the system is the number density correlation function. Defining the number density of monomers at a space point $x$

$$\rho(x) = \sum_i \delta(x - x_i),$$

the correlation function can be written as

$$\langle \rho(x)\rho(x')\rangle = \sum_i \langle \delta(x - x_i)\rangle \delta(x - x') + \sum_{i \neq j}\langle \delta(x - x_i)\delta(x' - x_j)\rangle.$$

The average density $\langle \rho(x)\rangle = \sum_i \langle \delta(x - x_i)\rangle = \sum_i p_i(x) = Np(x)$ is proportional to the probability $p(x)$ of finding any monomer of the chain at point $x$, while $\sum_{i \neq j}\langle \delta(x - x_i)\delta(x' - x_j)\rangle = N(N-1)p(x, x')$ is proportional to the joint probability of finding any monomer in $x$ and any other monomer in $x'$. This can be written as

$$p(x, x') = p(x'|x)p(x),$$

where the conditional probability can be written in terms of the radial distribution function $g(x, |x - x'|) = Np(x'|x)\langle \rho(x')\rangle^{-1}$, which has the advantage that can be calculated looping on the pairs of monomers of the chain instead that on the pairs of points of space. We obtain

$$\langle \rho(x)\rho(x')\rangle = \langle \rho(x)\rangle \delta(x - x') + \langle \rho(x)\rangle\langle \rho(x')\rangle g(x, |x - x'|) + o(1).$$

For a translationally-invariant system, like a model of an infinite fluid, the average density and the radial distribution function do not depend on the point $x$ and therefore the correlation function depends only on the distance $|x - x'|$. This is not the case for a small polymer. Taking advantage of the isotropy of the system, we selected a small region $\Omega$ around the center of mass of the chain and calculated

$$\tilde{g}(r) = \langle g(x, r)\rangle_{x \in \Omega} = \frac{\langle p(r|x)\rangle_{x \in \Omega}}{\tilde{\rho}(r)}$$

where $\tilde{\rho}(r)$ is the density of monomers at distance $r$ from the center of mass of the chain.

**Results**

The replica-exchange simulations of HIV-1-PRΔ96-99 and Brca1-304-394 are carried out in water and in 0.150M buffer in the temperature range from 300 K to 420 K (Table 1). Since the folding time of HIV-1-PRΔ96-99 is on the scale of seconds and we start the simulations from denatured conformations, no folding event is observed and we only sample its denatured state.

*The gyration radius is small in water and increases in buffer*

Unexpectedly, the distributions of the radii of gyration of HIV-1-PRΔ96-99 and Brca1-304-394 at 300K in both water and buffer are quite similar, even though they only share a similar length without any sequence similarity (Fig. 1). The two proteins in water show a peaked distribution of $R_g$ with averages 2.19 nm and 2.01 nm, respectively. In buffer, they develop a long tail towards higher values and both their averages increase to 2.52 nm. The dependence on the ionic strength is then more pronounced for the IDP, due to its more charged nature (Table S1 in the Supp. Mat.). The radius of gyration of the crystallographic HIV-1-PRΔ96-99 is 1.3 nm, so that the average size of the denatured state is approximately 50% larger than the native state in all solvent conditions.

An important reference length for assessing the compactness of the proteins is the radius of gyration of an ideal chain with the same length and the same local interactions (summarized by its persistence length $l_p$). The persistence length of the simulated proteins, calculated from the decorrelation of the directions of the bonds of the backbone, are 0.48±0.09 nm for HIV-1-PRΔ96-99 and 0.46±0.13 nm for Brca1-304-394. These values are slightly larger but comparable to those reported in the literature for proteins in general, i.e. 0.4 nm[25]. The mean radius of gyration of an ideal chain can be calculated as

$$R_g = \frac{l_K}{6^{1/2}} \left(\frac{L}{l_K}\right)^{1/2}, \tag{9}$$

where $l_K = 2l_p$ is the Kuhn length[13] and $L$ is the contour length of the protein ($L \approx 3 \cdot 0.14\, N$ nm, where $N$ is the number of residues and 0.14 nm is the mean bond length for the backbone). The reference values of $R_g$ for the ideal chain at 300K are 2.53 nm for HIV-1-PRΔ96-99 and 2.42 nm for Brca1-304-394 (dot-dashed lines in Fig. 1).

The average $R_g$ values for the two proteins in water are well below those of the ideal chain, while in buffer they are approximately equal to it. Applying standard homopolymer theory to the proteins, these results would suggest that both proteins are globular in water and approach the θ-point on addition of salts; considering that polymers are subject to pretransitional swelling [13], it is difficult to conclude whether they could reach the θ-point in buffer.

Exploring the behavior of the systems at different temperatures, we noticed a discrepancy between Amber and Calvados (Fig. S6 in the Supp. Mat.). Trying to investigate the motivations behind this, we noticed that the training of the latter was performed between 300 and 320 K, thus not guaranteeing its good performances outside such temperature range.

The average distance $\langle d(l) \rangle$ between residues at a linear distance $l$ along the chain does not follow the scaling $N^{3/5}$ expected for a random coil, the scaling $N^{1/2}$ of ideal chains or the constant behavior of coils (Fig. S7 in the Supp. Mat). On the contrary, it shows a monotonic sub-power-law shape for all simulations and a noisy end at $l \approx N$ due to poorer statistics.

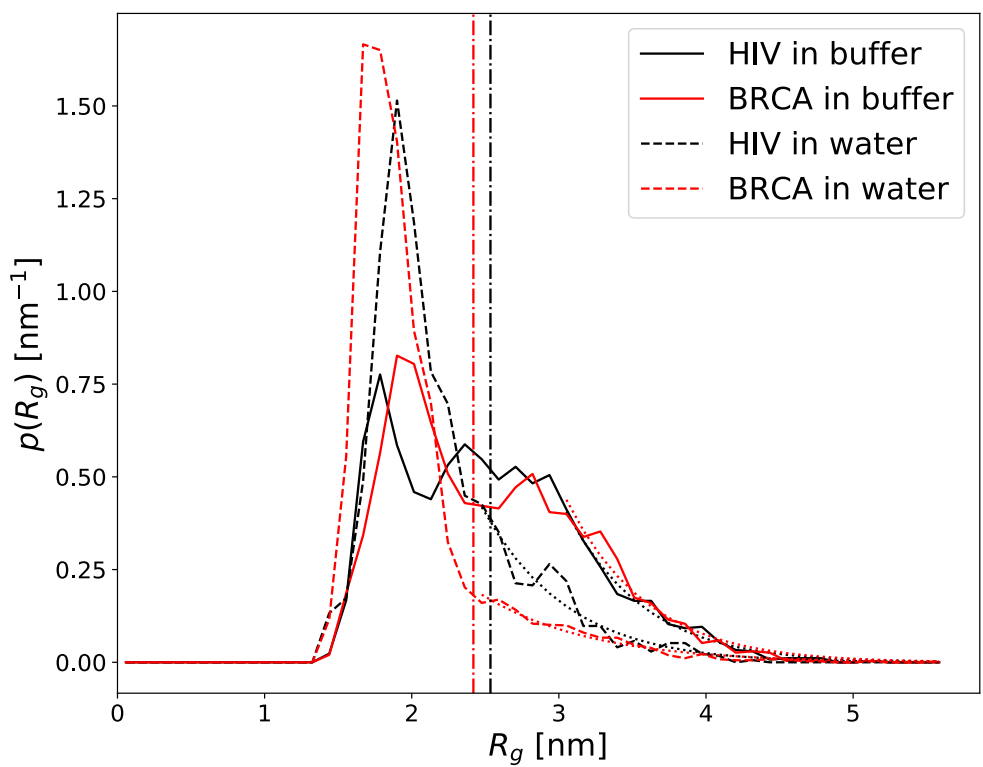

Figure 1: The distribution of the radius of gyration for HIV-1-PRΔ96-99 (solid black curve) and Brca1-304-394 (solid red curve), with averages 2.52 and 2.53 nm, respectively. Dashed lines indicate the results of simulations of the same proteins in water, with averages 2.19 and 2.01 nm, respectively. Dotted lines indicate fits of the tails of the distributions. The colored vertical bars indicate the radius of gyration for the ideal behavior.

*Hydrophobic collapse*

The median of solvent-accessible surface area (SASA) of hydrophobic residues for HIV-1-PRΔ96-99 in buffer (Fig. 2) is almost halfway between that of the native conformation and that of the maximum allowable accessibility[26], with hydrophobic residues being slightly more protected in water than in buffer. This is compatible to the degree of protection found in the intermediate state of proteins, where the collapse of hydrophobic residues can be experimentally quantified by hydrogen-deuterium exchange [27].

Also in the case of Brca1-304-394 in buffer, the hydrophobic SASA is consistently lower than in the fully extended conformation and the median is about 75% of that. Although the number of hydrophobic residues is much lower than in a structured protein, some protection can also be observed for IDPs, mainly due to local structuring of the chain (see below).

Both proteins have a slightly lower SASA in water than in buffer, and the fluctuations in SASA are also smaller in water.

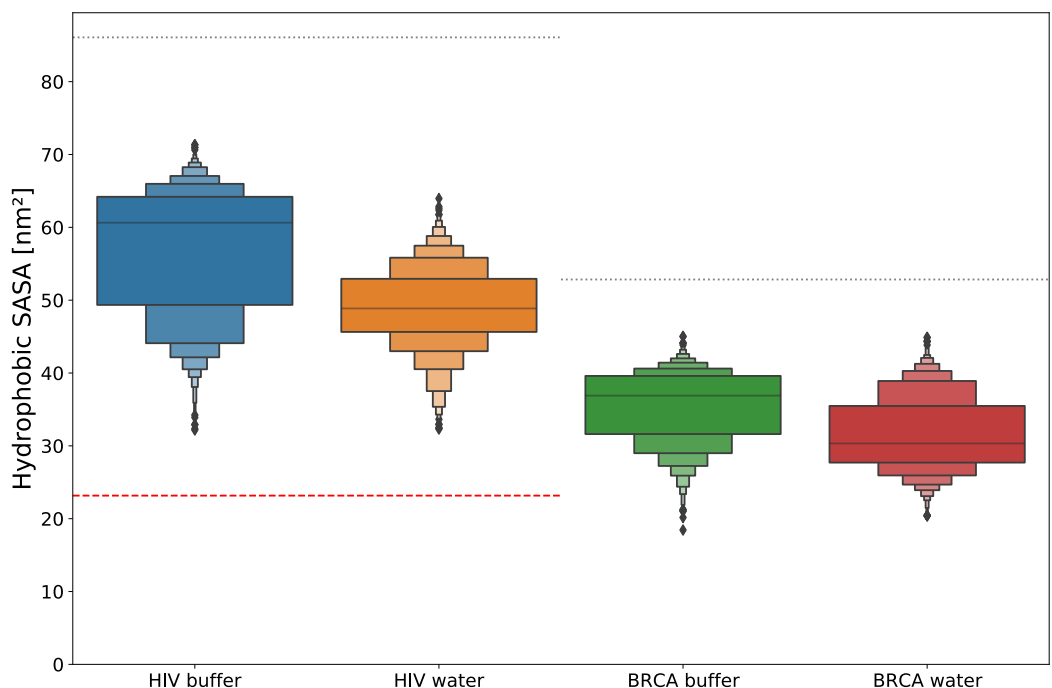

Figure 2: The distribution of hydrophobic SASA represented as a boxplot. The solid horizontal line in the central box is the median. The black dotted line is the maximum SASA that can be achieved for each protein [26], while the red dashed line is that of the native conformation.

*The density correlations in water do not agree with a globular phase*
Although the radii of gyration of the two proteins in water are smaller than that of an ideal chain and a hydrophobic collapse has occurred, suggesting that they are in the globular phase, the shapes of their density profile are more similar to those of a coil. We calculated the average density as a function of the distance $r$ from the center of mass, normalized to the radius of gyration, for the two proteins and compared them with that of an ideal chain and of a homopolymeric globule of the same length (Fig. 3a). HIV-1-PRΔ96-99 and Brca1-304-394 have profiles that are similar to each other and more similar to the monotonic decay of an ideal. In particular, they do not show the broad plateau of globules, which only begins decaying as the radius of gyration is approached. The density of the two proteins in buffer has a decay which is indistinguishable from that of an ideal chain, except at small values of $r$, which depends on the specific atomic structure of the amino acids and which cannot be recapitulated by non-interacting monomers.

The profound physical difference between the coil and the globular phases in a homopolymer is not so much in the size as in the density correlation function and its fluctuations[15]. In globules, the correlation length is smuch maller than the size of the globule and the density fluctuations are much smaller than the average; in coils, the correlation length is comparable to the size of the molecule and the density fluctuations are comparable to their average.

The relative two-point correlation function $\tilde{g}$ of the number density (calculated as described in the *Methodology* Section) is quite similar for both proteins in each solvent condition (Fig. 3b). It has some degree of structure, corresponding to oscillations associated with the excluded volume of the monomers, modulated by an overall decrease to 1. Correlations reach distances of around half of the radius of gyration for both solvent conditions. This behavior, except for the oscillations, is identical to that of the ideal chain. Vice versa, a homopolymeric globule shows a correlation function that reaches 1 already at 20% of the radius of gyration.

The relative density fluctuations, calculated in shells of increasing radius around the center of mass of the polymer (Fig. 3c), are also similar between the two proteins in each solvent condition. They are greater than 1 for small r, because the average number of monomers is small (because the volume is small) and therefore the density is very sensitive to the motion of the chain, and then decrease to about 0.5. It increases again for large r because the number of monomers is again small, now because the density is small.

The ideal chain has a similar shape, reaching a lower minimum with relative fluctuations of 0.3. The globule has a different shape, with a large plateau corresponding to relative fluctuations less than 0.1. The relative fluctuations of the two proteins in water are comparable to those of the ideal chain, while in buffer are even larger.

Summing up, the statistical features of the density of the two proteins both in water and in buffer are more similar to those of an ideal chain than those of a globule.

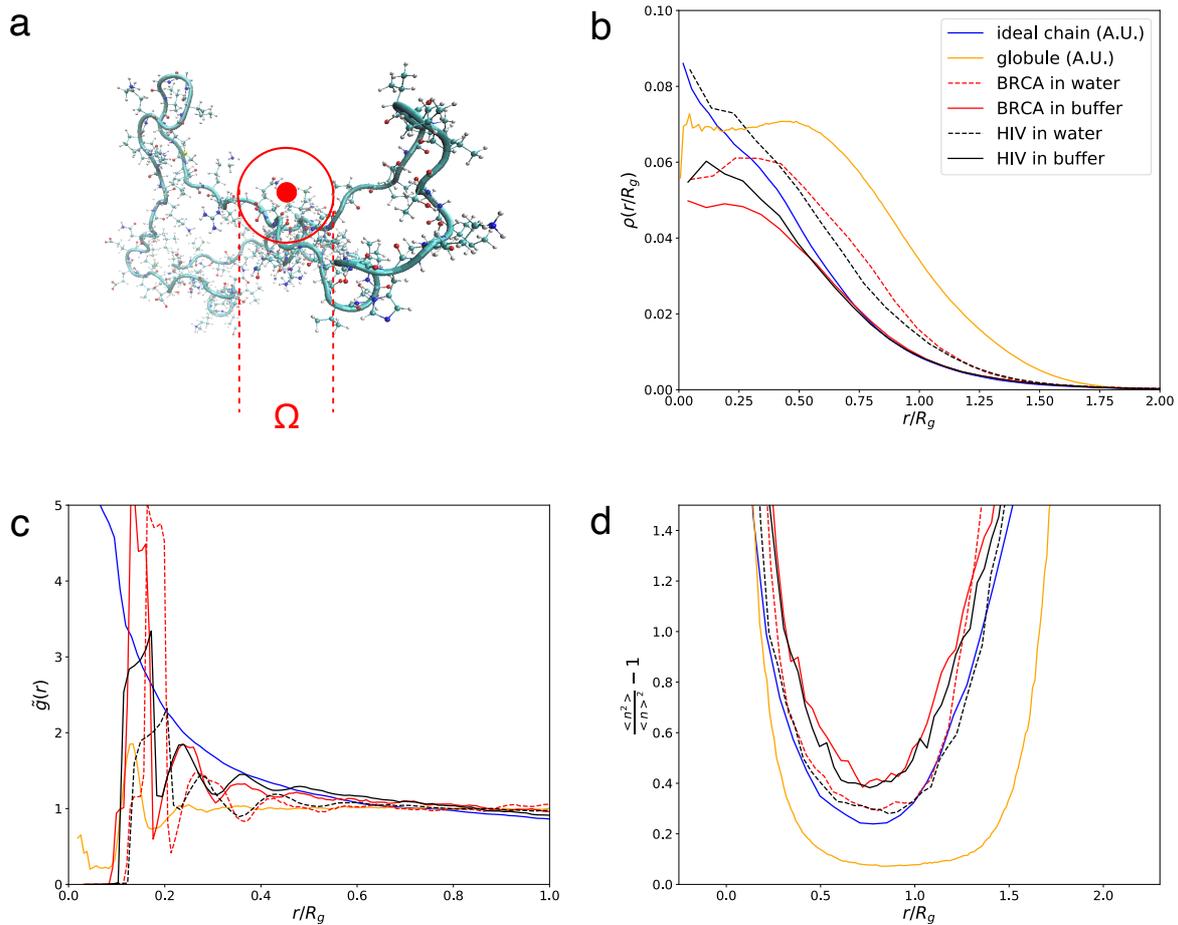

Figure 3: (a) A sketch of the for HIV-1-PRΔ96-99 deanatured monomer, where the center of mass and the spherical shell used to calculate the density are indicated in red. (b) The average density as a function of the distance $r$ from the center of mass of the system, normalized to the gyration radius $R_g$, for HIV-1-PRΔ96-99 (black curve) and Brca1-304-394 (red curve), and for a reference ideal chain (blue curve) and a homopolymeric globule (orange curve). (c) The normalized correlation function. (d) The density fluctuations in the spherical cells around the center of mass.

*Segments of the protein have different radii of gyration*

Another observation that distinguishes the proteins from homopolymers is the heterogeneity in the size of the different segments of the protein. We calculated the relative radius of gyration of all consecutive segments of length k with respect to the radius of gyration of the whole system (Fig. 4). Both proteins, and HIV-1-PRΔ96-99 in particular, show significant fluctuations in the radius of gyration between segments with lengths on the scale of less than 10 amino acids. Homopolymeric globules (dashed line in Figure 4) do not show this effect. Also in water, where the two proteins are even more compact than in buffer, the radius of gyration of the different segments is very variable (Fig. S8 in the Supplementary Material).

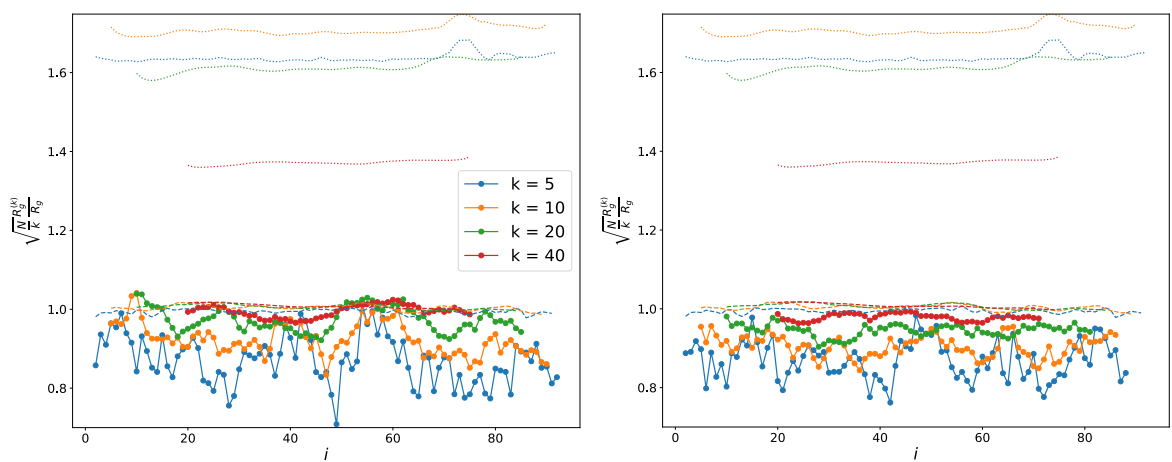

Figure 4: The mean radius of gyration calculated for all segments of the chain of length *k* in HIV-1-PRΔ96-99 (left panel) and Brca1-304-394 (right panel) in buffer. The dashed lines are the results of the same calculation for an ideal chain while the dotted lines are the result for a homopolymeric globule.

*Both proteins show residual structure*
The simulation of HIV-1-PRΔ96-99 reveals significant, albeit unstable, secondary structures at 300K even in buffer, where the protein is expected to behave as an ideal chain (upper-right panel in Fig. 5). Alpha helices are observed throughout the chain with a probability of between 10% and 20%. A native beta-hairpin is present in the middle of the chain with a probability of around 20%, and other small beta structures can be seen with probabilities below 10%. The patterns of secondary structure are similar in water, but their stability is approximately twice that in buffer (Fig. S9 in the Supp. Mat.)
Tertiary contacts are defined between residues whose atoms have a mutual distance of less than 0.4 nm and are separated by at least 4 other residues along the chain. They are localized in specific parts of the contact map (upper-left panel in Fig. 5) and have a typical frequency of about 5%.
Brca1-304-394 have an overall lower amount of both secondary and tertiary structure (lower panels in Fig. 5), which increases in water (Fig. S9 in the Supp. Mat.). With the exception of one C-terminal alpha-helix, the structure is more uniform along the chain and its amount is less than 5%.

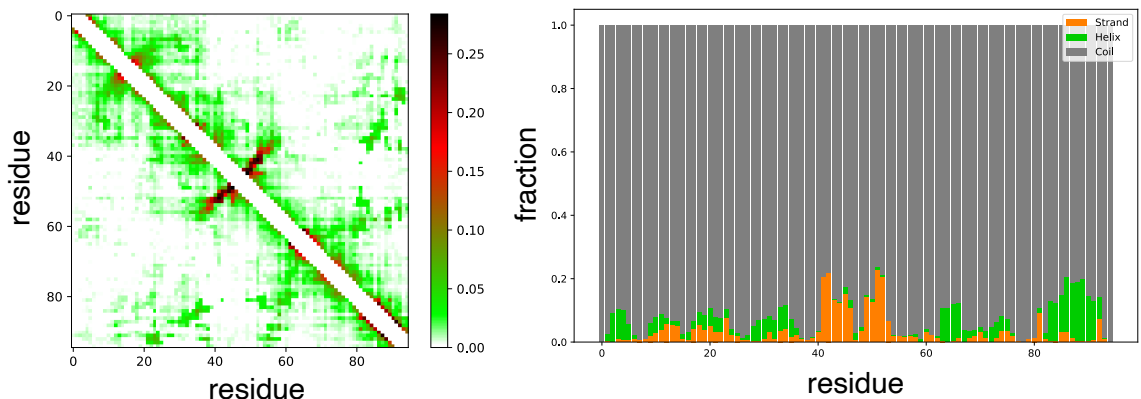

**HIV-1-PR-Δ96-99**

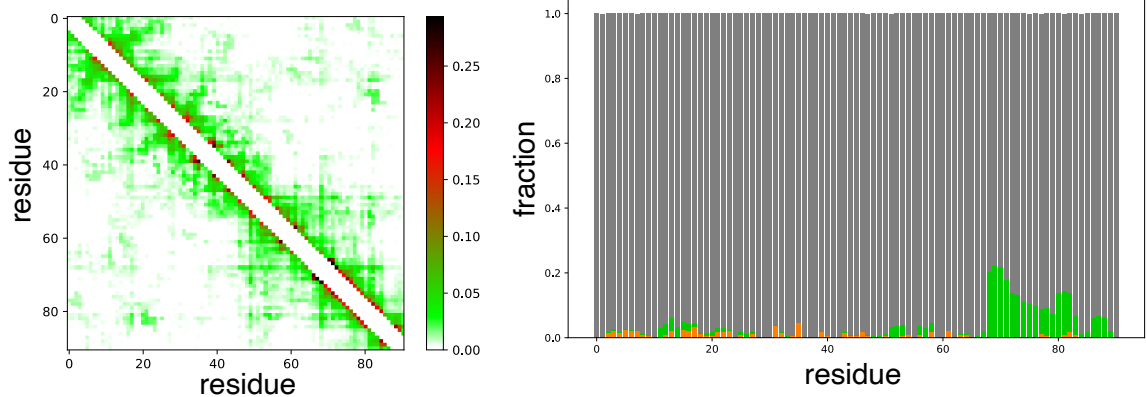

**Brca1-304-394**

Figure 5: The residual tertiary (left panels) and secondary (right panels) structure of HIV-1-PRΔ96-99 and Brca1-304-394 in buffer.

**Discussion and Conclusions**

Following the development of force fields that can realistically describe disordered proteins, molecular dynamics simulations have become a powerful tool for describing the metastable denatured state of structured proteins and comparing it with the ensemble of conformations populated by IDPs. The main computational limitation to this goal is the need for a large box to contain the disordered proteins, together with a large number of solvent molecules whose coordinates must be integrated at each time step of the simulation. A poor choice of box size would lead to spurious interactions between regions of the chain across the periodic boundary condition and thus to artefactual results.

We developed a multiscale approach to speed up the simulations by limiting the swelling of the chain with a constraining potential, combining the result of the all-atom simulations with that of a coarse-grained model, and reweighting the sampled conformations a posteriori. In this way, the more compact conformations, which are expected to depend substantially on the detailed atom-atom interaction, are treated with the Amber force field; the more extended conformations, which are mainly stabilized by entropy, are treated with the coarse-grained Calvados model.

The two models give compatible results around 300K, but not above 320K. This is not unexpected since Calvados is based on a data-driven potential trained on experimental data obtained at biological temperatures. This discrepancy prevents us from studying in detail the coil-globule transition of the protein chains that is sometimes observed within the denatured state of proteins[5].

The simulations show that the radius of gyration of the two proteins in water is significantly smaller than that of a corresponding ideal chain, while in buffer it is approximately that of an ideal chain. In the language of homopolymer theory, this would place them in the globule phase and exactly at the θ-point, respectively. The globular phase of HIV-1-PRΔ96-99 is associated with a significant hydrophobic collapse in both solvent conditions, which is much more limited for Brca1-304-394.

The strong dependence of the radius of gyration on the ionic strength of the solution means that disordered proteins must be simulated in a realistic buffer to obtain meaningful results. At the same time, it provides us with ensembles of conformations with different size distributions that can be analyzed in detail.

The size distribution of the two proteins we simulated in water suggests a classification as globules. However, the density profile, the density-density correlation function and the density fluctuations show the behaviour expected for an ideal chain or for a coil in both solvent conditions. The picture that emerges is then inconsistent from the perspective of homopolymer theory. If one wants to force the interpretation of the data within the standard theory, it is the change in density correlations, not the size, that marks the coil-globule transition in a homopolymer[15]. In any case, the mismatch between size and correlation properties in our simulations suggests that homopolymer theory falls short in describing the polymeric phase of disordered proteins.

In fact, proteins are not homopolymers. The sequence-dependent nature of the size of the protein can be made explicit by studying the radius of gyration of short chain segments, which is very variable and fluctuates much more than in homopolymers. The smoking gun that the concepts borrowed from homopolymer physics cannot be directly applied to the disordered states of proteins is the presence of transient secondary and tertiary structure. Although highly unstable, the secondary structure we observed in HIV-1-PRΔ96-99 and in Brca1-304-394 effectively makes the chain shorter than its nominal size. This effect can be accounted for in the persistence length of the polymer and we expect that it does not significantly alter the phases of the polymer. The specific tertiary structure that we observe in the simulation of both proteins is something that is not present in homopolymers and can greatly influence the size of the chain.

Even the standard theory of random heteropolymers[28], which describes the heterogeneous nature of the interaction between the amino acids but can only be worked out at the level of the mean field, cannot account for the discrepancy observed in the simulations. In fact, focusing on the average density as an order parameter and averaging out the sequence specificity, it predicts a standard coil-globule transition at a temperature renormalized by the heterogeneity of the interactions with respect to the homopolymer.

It is difficult to build a theory to quantitatively describe the effect of the transient structure on the size of the protein. In fact, one consequence of this sequence-dependent structure is that it violates the scaling laws (Fig. S7) that make the analysis of homopolymers particularly straightforward. We can only observe qualitatively that the presence of residual structure effectively shortens the protein chain and thus reduces the radius of gyration. Thus, we expect that the heterogeneous nature of protein interactions will always act to make the protein more globular than a homopolymeric theory would suggest.

Experimental support for our findings is provided by the fact that comparing the average size of proteins obtained by different experiments, such as SAXS and FRET, gives different results[29], suggesting that the size of the protein is not a good order parameter to identify its polymeric phase. In particular, it was shown[5,11,12] that single-molecule FRET experiments are compatible with a compact state as the concentration of chemical denaturant is reduced for 14 structured proteins, while SAXS reports extended structures under the same conditions. These results could not been explained in a homopolymeric picture.

However, if we need to classify the denatured state of a protein according to the standard coil-globule concepts, we should follow a criterion based on density correlations and not on the size of the polymer. In this case, HIV-1-PRΔ96-99 and in Brca1-304-394 would be in a coil phase in all conditions of ionic strength, even if the radius of gyration is small. Indeed, a coil-like state would confer to the protein a kinetic advantage with respect to a dense globule[30].

We chose these two proteins because they are of similar length, typical of medium-sized proteins but short enough to allow us to simulate them, and they have the standard amino acid composition of structured proteins and IDPs, respectively. Although the specific transient structure that we observed to break the homopolymeric interpretation of the states of these proteins depends on their sequence, we believe that this is a general conclusion that applies to most proteins.

**Supplementary Material**

The Supplementary Material contains some additional figures.

**Acknowledgments**

We acknowledge the computational support of CINECA through the INFN project Biophys.

**Conflict of interest**

The authors declare no conflicts of interest.

**Data availability**

The trajectories generated by Gromacs at 300K can be downloaded at https://doi.org/10.13130/RD_UNIMI/8HSLTI

**Author contributions**

FR and GP performed the calculatios; FR, RC and GT analyzed the results; FR and GT conceived the project and wrote the manuscript.

**References**

[1] K.A. Dill, and D. Shortle, "Denatured states of proteins," Annu Rev Biochem **60**, 795–825 (1991).
[2] V.N. Uversky, "Natively unfolded proteins: A point where biology waits for physics," Protein Science **11**(4), 739–756 (2002).
[3] B. Robson, and R.H. Pain, "Analysis of the Code Relating Sequence to Conformation in Proteins: Possible Implications for the Mechanism of Formation of Helical Regions," J. Mol. Biol **58**, 237–259 (1971).
[4] D. Shortle, and M.S. Ackerman, "Persistence of native-like topology in a denatured protein in 8 M urea," Science (1979) **293**(5529), 487–489 (2001).
[5] E. Sherman, and G. Haran, "Coil-globule transition in the denatured state of a small protein.," Proc Natl Acad Sci U S A **103**(31), 11539–11543 (2006).
[6] V.N. Uversky, J.R. Gillespie, and A.L. Fink, "Why are 'natively unfolded' proteins unstructured under physiologic conditions?," Proteins: Structure, Function and Genetics **41**(3), 415–427 (2000).

[7] C. Tanford, K. Kawahara, and S. Lapanje, "Proteins in 6 m Guanidine Hydrochloride: demonstration of random coil behavior," Journal of Biological Chemistry **241**(8), 1921–1923 (1966).

[8] J.E. Kohn, I.S. Millett, J. Jacob, B. Zagrovic, T.M. Dillon, N. Cingel, R.S. Dothager, S. Seifert, P. Thiyagarajan, T.R. Sosnick, M.Z. Hasan, V.S. Pande, I. Ruczinski, S. Doniach, and K.W. Plaxco, "Random-coil behavior and the dimensions of chemically unfolded proteins," Proceedings of the National Academy of Sciences **101**(34), 12491–12496 (2004).

[9] J.M. Flanagan, M. Kataoka, D. Shortle, and D.M. Engelman, "Truncated staphylococcal nuclease is compact but disordered.," Proceedings of the National Academy of Sciences **89**(2), 748–752 (1992).

[10] J.A. Marsh, and J.D. Forman-Kay, "Structure and Disorder in an Unfolded State under Nondenaturing Conditions from Ensemble Models Consistent with a Large Number of Experimental Restraints," J Mol Biol **391**(2), 359–374 (2009).

[11] Y. Hayashi, N. Miura, N. Shinyashiki, S. Yagihara, and S. Mashimo, "Globule-Coil Transition of Denatured Globular Protein Investigated by a Microwave Dielectric Technique," Biopoly **54**, 388–397 (2000).

[12] K.M. Fiebig, H. Schwalbe, M. Buck, L.J. Smith, and C.M. Dobson, "Toward a description of the conformations of denatured states of proteins. Comparison of a random coil model with NMR measurements," Journal of Physical Chemistry **100**(7), 2661–2666 (1996).

[13] A.Y. Grosberg, and A. Khokhlov, *Statistical Physics of Macromolecules* (AIP Press, Woodbury, NY, 1994).

[14] N. Fitzkee, and G. Rose, "Reassessing random-coil statistics in unfolded proteins," Proceedings of the National Academy of Sciences **101**(34), 12497–12502 (2004).

[15] I.M. Lifshits, A. Yu Grosberg, and A.R. Khokhlov, "Volume interactions in the statistical physics of a polymer macromolecule," Sov. Phys. Usp **22**, 123 (1979).

[16] C.D. Sfatos, A.M. Gutin, and E.I. Shakhnovich, "Phase diagram of random copolymers," Phys Rev E **48**(1), 465–475 (1993).

[17] P. Robustelli, S. Piana, and D.E. Shaw, "Developing a molecular dynamics force field for both folded and disordered protein states," Proc Natl Acad Sci U S A **115**(21), E4758–E4766 (2018).

[18] G. Tesei, A.I. Trolle, N. Jonsson, J. Betz, F.E. Knudsen, F. Pesce, K.E. Johansson, and K. Lindorff-Larsen, "Conformational ensembles of the human intrinsically disordered proteome," Nature, (2024).

[19] H.I. Rösner, M. Caldarini, A. Prestel, M.A. Vanoni, R.A. Broglia, A. Aliverti, G. Tiana, and B.B. Kragelund, "Cold Denaturation of the HIV-1 Protease Monomer," Biochemistry **56**(8), 1029–1032 (2017).

[20] S. Müller-Späth, A. Soranno, V. Hirschfeld, H. Hofmann, S. Rüegger, L. Reymond, D. Nettels, and B. Schuler, "Charge interactions can dominate the dimensions of intrinsically disordered proteins," Proceedings of the National Academy of Sciences **107**(33), 14609–14614 (2010).

[21] H.I. Rösner, M. Caldarini, G. Potel, D. Malmodin, M.A. Vanoni, A. Aliverti, R.A. Broglia, B.B. Kragelund, and G. Tiana, "The denatured state of HIV-1 protease under native conditions," Proteins: Structure, Function and Bioinformatics **90**(1), 96–109 (2022).

[22] W.L. Jorgensen, J. Chandrasekhar, J.D. Madura, R.W. Impey, and M.L. Klein, "Comparison of simple potential functions for simulating liquid water," J Chem Phys **79**(2), 926–935 (1983).

[23] D. Van Der Spoel, E. Lindahl, B. Hess, G. Groenhof, A.E. Mark, and H.J.C. Berendsen, "GROMACS: fast, flexible, and free," J Comput Chem **26**(16), 1701–1718 (2005).

[24] G.A. Tribello, M. Bonomi, D. Branduardi, C. Camilloni, and G. Bussi, "PLUMED 2: New feathers for an old bird," Comput Phys Commun **185**(2), 604–613 (2014).

[25] H.X. Zhou, "Polymer Models of Protein Stability, Folding, and Interactions," Biochemistry **43**(8), 2141–2154 (2004).

[26] M.Z. Tien, A.G. Meyer, D.K. Sydykova, S.J. Spielman, and C.O. Wilke, "Maximum Allowed Solvent Accessibilites of Residues in Proteins," PLoS One **8**(11), e80635 (2013).

[27] W.A. Houry, and H.A. Scheraga, "Structure of a Hydrophobically Collapsed Intermediate on the Conformational Folding Pathway of Ribonuclease A Probed by Hydrogen−Deuterium Exchange," Biochemistry **35**(36), 11734–11746 (1996).

[28] E.I. Shakhnovich, and A.M. Gutin, "Formation of unique structure in polypeptide chains. Theoretical investigation with the aid of a replica approach," Biophys Chem **34**(3), 187–199 (1989).

[29] G. Haran, "How, when and why proteins collapse: The relation to folding," Curr Opin Struct Biol **22**(1), 14–20 (2012).

[30] A.M. Gutin, V.I. Abkevich, and E.I. Shakhnovich, "Is Burst Hydrophobic Collapse Necessary for Protein Folding?," Biochemistry **34**(9), 3066–3076 (1995).